\documentclass[twocolumn,prl,showpacs,preprintnumbers,amsmath,amssymb]{revtex4}

\usepackage[dvips]{graphicx}
\usepackage{dcolumn}
\usepackage{bm}

\begin{document}

\preprint{final version}

\title{The Loss of Multifractality in Migraines}

\author{Miroslaw Latka}
\email{mirek@if.pwr.wroc.pl}
\homepage{http://www.if.pwr.wroc.pl/~mirek}
\affiliation{
Institute of Physics, Wroclaw University of Technology, Wybrzeze Wyspianskiego 27,
                    50-370 Wroclaw, Poland\\
}%

\author{Marta Glaubic-Latka}
\affiliation{Opole Regional Medical Center
 Al. Witosa 26, 45-401 Opole, Poland 
}\email{mlatka@wcm.opole.pl}
\author{Dariusz Latka}
 \affiliation{Department of Neurosurgery, Opole Regional Medical Center,
              Al. Witosa 26, 45-401 Opole, Poland 
}
 \email{dlatka@wcm.opole.pl}

\author{Bruce J. West}
\email{westb@aro-emh1.army.mil}
\affiliation{Mathematics Division, Army Research Office, P.O. Box 12211, Research Triangle, NC 27709-2211, USA
}

\date{April 2, 2002}

\begin{abstract}
We study the middle cerebral artery blood flow velocity (MCAfv) in humans
using transcranial Doppler ultrasonography (TCD). The time series of the
axial flow velocity averaged over a cardiac beat interval is found to
exhibit clear multifractal properties for healthy subjects. We observe a
loss of multifractality for subjects with migraine.

\end{abstract}

\maketitle

Migraine headaches have been the bane of humanity for centuries, afflicting
such notables as Ceasar, Pascal, Kant, Beethoven, Chopin, and Napoleon.
However, its aetiology and pathomechanism have not to date been
satisfactorily elucidated. Herein we demonstrate that the characteristics of
the time series associated with cerebral blood flow (CBF) significantly  differs 
between that of normal healthy individuals
and  migraineurs. Migraine is considered by many to be a functional
neurological disorder in which \textit{CBF autoregulation failure} may be
responsible for the onset of symptoms. The results of our data analysis
support this interpretation.

Physiological signals, such as CBF time series, are typically generated by
complex self-regulatory systems that process inputs with broad range of
accessible values. Even though this type of time series may fluctuate in an
irregular and complex manner, they frequently exhibit \textit{self-affine}
or \textit{fractal} properties which can be characterized by a single global
parameter -- the fractal dimension $D$ or equivalently the Hurst exponent $H$ 
($D=2-H$)\cite{bass94}.
The salient property of mathematical random fractal process is the existence of
long-range correlations for $H\neq 1/2$. The studies of the cardiac
beat-to-beat variability have shown the existence of strong long-range
correlations in healthy subjects and demonstrated the breakdown of
correlations in disease \cite{peng93} (see also \cite{goldberger99} and
references therein). A similar pattern was observed in fluctuations in the
stride interval in human gait. The strength of correlations was
significantly reduced both by aging and a neurodegenerative disease. This
effect is frequently referred to as the loss of complexity \cite
{hausdorff95,hausdorff97,west98a,walleczek2000}. Complexity decreases with
the convergence of the Hurst exponent on $H=1/2$.

While the properties of monofractals are determined by the global Hurst 
exponent there exists a more general class of heterogenous signals known as 
\textit{multifractals} which are made up of many interwoven subsets with
different \textit{local} Hurst exponent $h$. The statistical properties of
these subsets may be characterized by the distribution of fractal dimensions $D(h)$.
Ivanov \textit{et al} \cite{ivanov99} established that the healthy human heartbeat
interval exhibits multifractal properties and uncovered the loss of
multifractality for the life-threatening condition known as congestive heart
failure. Anticipating our result, herein we find the loss of multifractality 
in the time series of middle cerebral artery blood flow velocity is almost 
entirely lost in subjects with "severe"  migraine even in headache-free intervals.

A healthy human brain is perfused by blood flowing laminarly through the
cerebral vessels providing brain tissue with substrates,
such as oxygen and glucose. It turns out that CBF is relatively stable with typical values between 45 and
65 ml/100g of brain tissue per second, despite variations in systemic
pressure as large as 100 $Torr$. This phenomenon is known as \textit{%
cerebral autoregulation} and has been thoroughly documented not only in
humans but also in animals \cite{heistad83,paulson90}. Changes in
cerebrovascular resistance (CVR) of small precapillary brain arteries is a
major mechanism responsible for maintaining relatively constant cerebral
blood flow. For example, CVR may increase due to mechanoreceptor constrictions of these
arteries caused by the elevation of intracranial pressure (ICP) and/or
biochemically mediated constrictions associated with the decrease of CO$_{2}$
arterial content. It is worth pointing out that changes in \textit{local}
cerebral activity may affect the regional cerebral flow (rCBF). The increase
in metabolism leads to the local drop of extracellular pH (which is
associated with the elevated production of CO$_{2}$, lactic acid, and other
metabolites) which in turn enhances the rCBF.

These complex cerebral flow autoregulation mechanisms are supposed to
be influenced or even to be fundamentally altered in many pathological states 
\cite{olesen90}. However, despite the significant advances in brain
diagnostic imaging techniques many functional aspects of CBF regulation are
not fully understood. For example, migraine -- prevalent, hemicranial
(asymmetric) headache is among the least understood diseases. The leading
neurovascular theory identifies serotonin, a strong vasculomotoric agent, as
the main biochemical factor. In the last several years, CBF in migraine
patients has been studied using diverse techniques such as Xenon-133 uptake
measurements, cerebral termography (CT), single proton emission tomography
(SPECT) and positron emission tomography (PET). Although the pathophysiology
of migraine headache has not been unequivocally explained, some experimental
data reveal clear interhemispheric blood flow asymmetry in some parts of
brain of migraineurs even during headache-free intervals \cite
{battistella90, mirza98, shyhoj89}.

Transcranial Doppler ultrasonography enables high-resolution measurement of
MCA flow velocity. Even though this technique does not allow us to directly
determine CBF values, it may help to elucidate the nature and role of
vascular abnormalities associated with migraine. Some previous studies have
discovered  significant changes in cerebrovascular reactivity in migraine
patients \cite{heckmann98}. In this work we look for the signature of the
migraine pathology in the dynamics of cerebral autoregulation. In
particular, we investigate the fractal and multifractal properties of the
human MCAfv time series.

The dynamical aspects of the cerebral autoregulation were recognized by
Zhang \textit{et al.} \cite{zhang98}. Keunen \textit{et al.} \cite{keunen94,
keunen96} applied the attractor reconstruction technique along with the
Grassberger-Procaccia algorithm and the concept of surrogate data to look
for the manifestations of the nonlinear dynamics in continuous waveforms of
TCD signals. Rossitti and Stephensen \cite{rossitti94} used the relative
dispersion of the MCAfv velocity time series to reveal its fractal nature.
West \textit{et al.} \cite{west99a} extended this line of research by taking
into account the more general renormalization-group properties of fractal
time series. Both studies \cite{rossitti94,west99a} showed that the
beat-to-beat variability in the flow velocity has a long-time memory and is
persistent with the average value of the Hurst exponent $H=0.85\pm 0.04$,
a value consistent with that found earlier for interbeat interval 
time series of the human heart.

We measured MCAfv using the Multidop T DWL Elektronische Systeme
ultrasonograph. The 2-MHz Doppler probes were placed over the temporal
windows and fixed at a constant angle and position. The measurements were
taken continuously for approximately two hours in the subjects at supine
rest. The study comprised eight migraineurs (who used to experience 1-3 migraines per month) 
and five healthy individuals. The migraineur group underwent the standard diagnostic procedures e.g. computer
tomography, EEG) to exclude illnesses not associated with cerebral
autoregulation. An example of a typical measured MCAfv time series is shown
in Figure 1 for the first thousand of the recorded beats of the subject's
heart. The total time series has over eight thousand data points for 
a two hour data record.

Successive increments of mathematical fractal random processes are
independent of the time step. They are correlated with the coefficient of
correlation $\rho $ which is determined by the formula $2^{2H}=2+2\rho $.
Thus for $H\neq 1/2$ there exist long-range correlations, that is, $\rho
\neq 0$. It turns out the Hurst exponent also determines the \textit{scaling}
properties of the fractal time series. If $y(t)$ is a fractal process with
Hurst exponent $H,$ then $y_{c}=y(ct)/c^{H}$ is another fractal process with
the same statistics. This type of scaling is called renormalization. The
variance of self-affine time series is proportional to $\Delta t^{2H}$ where 
$\Delta t$ is the time interval between measurements. A number of algorithms
which are commonly used to calculate the Hurst exponent are based on this
property.

Herein we employ the detrended fluctuation analysis (DFA) introduced into
the study of biomedical time series by Peng \textit{et al.} \cite{peng94}. 
Let ${\{v_{i}\}}_{i=1}^{N}$ be the experimental time series of the middle
cerebral artery blood flow velocity (MCAfv) $v$. First, the time series is
aggregated: $y(k)=\sum_{i=1}^{k}(v_{i}-\bar{v}),k=1,..,N$, where $\bar{v}$
is the average velocity. Then, for segments of the aggregated time series of
length $n$ the following quantity is calculated: 
\begin{equation}
F(n)=\overline{\sqrt{{{\frac{1}{{N}}}%
\sum_{k=n_{0}}^{n+n_{0}}[y(k)-y_{n_{0}}(k)]^{2}}}},  \label{F}
\end{equation}
where $y_{n_{0}}$ is a least square line fit to the data segment which
starts at $n_{0}$ and ends at $n+n_{0}$. The bar in the above equation
denotes an average over all possible starting points $n_{0}$ of data
segments of length $n$. Thus, for a given data box size $n$, $F(n)$ gives
the characteristic size of fluctuations of the aggregated and detrended time
series. If the aggregated time-series is fractal then $F(n)\sim n^{H},$ so
one obtains the Hurst exponent from a linear least-square fit to $F(n)$ on
double log graph paper. However, West has emphasized \cite{bass94, west99a}
the importance of possible periodic modulations of quantities such as $F(n)$%
. These modulations are intimately related to the renormalization-group
properties of fractal time series and may be accounted for with the help of
the following fit function: 
\begin{equation}
F_{X}(n)=n^{H}\exp [\alpha +\lambda \cos (\gamma \ln n)].
\label{fitFunction}
\end{equation}
Here again the Hurst exponent is determined by the slope of the fitting
curve, but now the curve also has a harmonic modulation in the logarithm of
the length $n$ of the data segment.

Fig. \ref{DFAFig} shows the typical DFA analysis for a healthy subject.
The circles in this figure are the calculated values of $F(n)$ and the solid
line is the best renormalization group fit, \textit{cf.} Eqn. (\ref
{fitFunction}). The best-fit parameters are given in the inset.
 
The average Hurst exponent obtained from ten measurements 
of healthy subjects is equal to 0.85 and coincides with the value
reported previously \cite{rossitti94,west99a}. Suprisingly enough,
the DFA analysis of 14 measurements of migraineurs yielded the 
average value of $H$ equal to 0.83. While the observed difference 
is insignificant, we emphasize that most of the measuremenents were
performed in headache-free intervals. Our further studies will address
the question as to whether the Hurst exponent changes \textit{during}
the migraine episode. 

We have already pointed out that in order to describe the scaling properties
of multifractal signals it is necessary to use many local Hurst (or
H\"{o}lder) exponents. Formally, the H\"{o}lder exponent $h(x_0)$ of a
function $f$ at $x_0$ is defined as the largest exponent such that there
exists a polynomial $P_{n}(x)$ of order $n$ that satisfies the following
condition \cite{muzy91,muzy93a,muzy93b,mallat98}: 
\begin{equation}
\mid f(x)-P_{n}(x-x_0) \mid= O({\mid x-x_0 \mid}^h)  \label{Holder}
\end{equation}
for $x$ in a neighborhood of $x_0$. Thus the H\"{o}lder exponent measures
the singularity of a function at a given point. For example, $h(x_0)=1.5$
implies that the function $f$ is differentiable at $x_0$ but its derivative
is not. The singularity lies in the second derivative of $f$. The
singularity spectrum $D(h)$ of the signal may be defined as the function
that for a fixed value of $h$ yields the Hausdorff dimension of the set of
points $x$ where the exponent $h(x)$ is equal to $h$ \cite{lastwave}.

In Fig. \ref{comparisonD(h)} we compare the averaged singularity spectrum 
of the healthy subjects with that of migraineurs. It is apparent that the 
multifractal properties of migraineurs are significantly reduced which is
reflected by the vastly constricted interval for 
the distribution of fractal dimension $D(h)$.

It seems that the changes in the cerebral autoregulation associated with
migraine can modify the multifractality of MCA blood flow much more strongly
than monofractal properties characterized by the single global Hurst
exponent. The loss of multifractality may persist in some subjects 
even in pain-free intervals. The more detailed analysis of the clinical data will be presented
elsewhere.


\bibliography{doplerPRL}%

\begin{figure}
\includegraphics{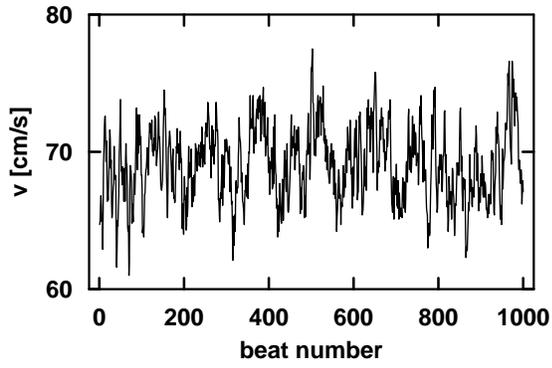}
\caption{MCAfv time series for a healthy subject.}
\label{KarolinaTrace}
\end{figure}

\begin{figure}
\includegraphics{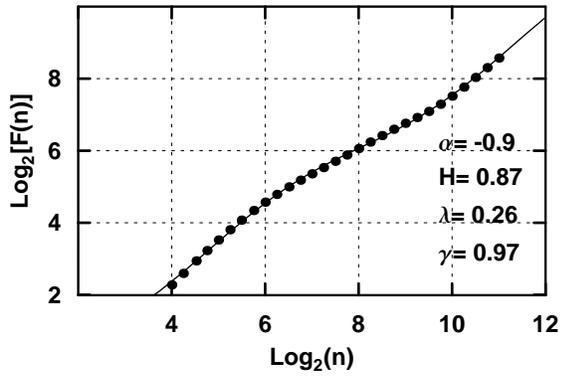}
\caption{DFA analysis of the MCAfv time series 
         shown in Fig. \ref{KarolinaTrace}}
\label{DFAFig}
\end{figure}

\begin{figure*}
\includegraphics{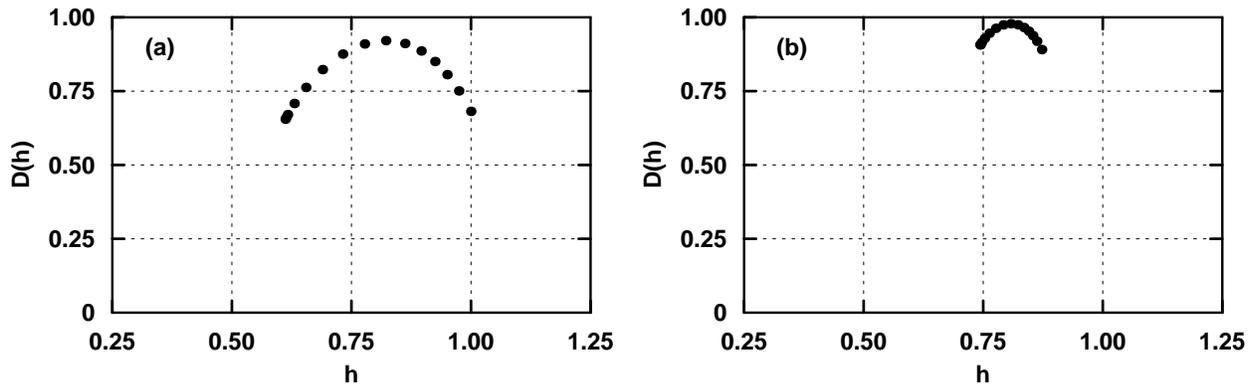}
\caption{Comparison of the averaged singularity spectrum $D(h)$ of the healthy
subjects (a) with that of the migraineurs (b). The spectra were  computed using the WTMM.
The analyzing wavelet was the second derivative of the Gaussian. The spectrum in  Fig. (a)
is the average of 10 measurements of five subjects. The spectrum in Fig. (b) is the average
of 14 measurements of eight subjects}
\label{comparisonD(h)}
\end{figure*}

\end{document}